\documentstyle[aps,prl,twocolumn,graphicx,floats]{revtex}
\begin{document}

\newcommand{\alex}[1]{({\bf\textsl{#1}})}
\newcommand{\andrei}[1]{({\bf\textsf{#1}})}

\preprint{\vbox{\hbox{IASSNS-AST-02/xx}},
  \vbox{hep-ph/0202095}}
\draft \wideabs{
\title{A new solution to the solar neutrino deficit}
\author{Alexander Friedland$^1$
%\footnote{New address:Theoretical
%    Division, T-8, MS B285, Los Alamos National Laboratory,
%    Los Alamos, NM 87545} 
and Andrei Gruzinov$^{2}$}
\address{
$^1$School of Natural Sciences, Institute for Advanced Study,
Princeton NJ, USA 08540\\
$^2$New York University, 4 Washington Place, New York, NY 10003 }

\date{November 27, 2002}
\maketitle

%\narrowtext

%\setcounter{footnote}{0} 
\setcounter{page}{1}
\setcounter{section}{0} \setcounter{subsection}{0}
\setcounter{subsubsection}{0}

%%%%%%%%%%%%%%%%%%%%%%%%%%%%%%%%%%%%%%%%%%%%%%%%%%%%%%%%%%%%%%%%%%%%%%%
\begin{abstract}
  We reexamine the transition magnetic moment solution to the solar
  neutrino problem. We argue that the absence of large time variations
  in the Super-Kamiokande rate provides strong evidence against
  spin-flavor flip in the solar convective zone.  Spin-flavor flip
  could, however, occur in the primordial magnetic field in the
  radiative zone. We compute the longest lived toroidal mode for this
  field and show that spin-flavor flip in the radiative zone can
  account for all available solar data.
\end{abstract}
}
%\draft

\section{Introduction}

In recent years, the progress in
the field of solar neutrino physics has been remarkable. This progress
culminated when the SNO collaboration conclusively demonstrated 
the presence of a non-electron neutrino component in the flux of
solar neutrinos \cite{SNOPRL,SNONC}.

The SNO experiment measured the flux of the $^8$B solar neutrinos
using three different reactions: (i) the charged current (CC) process
$\nu_e+d\rightarrow p+p+e^-$, (ii) the elastic scattering (ES) process
$\nu_x(\bar{\nu}_x)+e^-\rightarrow\nu_x(\bar{\nu}_x)+e^-$, and (iii)
the neutral current (NC) process $\nu_x(\bar{\nu}_x)+d\rightarrow
p+n+\nu_x(\bar{\nu}_x)$. A high statistics ES measurement, in the
same energy range, had also been previously carried out by the
Super-Kamiokande (SK) experiment \cite{SK2001}. While the CC process
selects only electron neutrinos, the other two processes are sensitive
to neutrinos \emph{as well as antineutrinos} of all active flavors.

The results of the measurements are as follows
\cite{SNONC,SK2002}:\footnote{$R_{\rm SNO}^{\rm NC1}$ was found
  assuming undistorted $^8$B neutrino energy spectrum, while $R_{\rm
    SNO}^{\rm NC2}$ was found when this assumption was relaxed.}
\begin{eqnarray}
  \label{eq:8Bexprates}
\label{eq:CC}
  R_{\rm SNO}^{\rm CC} &=& 
  1.76\pm 0.11\times 10^6 \mbox{cm}^{-2}\mbox{s}^{-1},\\
\label{eq:ESSNO}
  R_{\rm SNO}^{\rm ES} &=& 
  2.39\pm 0.27\times 10^6 \mbox{cm}^{-2}\mbox{s}^{-1},\\
\label{eq:ESSK}
  R_{\rm SK}^{\rm ES} &=&
  2.35\pm 0.08\times 10^6 \mbox{cm}^{-2}\mbox{s}^{-1},\\
\label{eq:NC1}
  R_{\rm SNO}^{\rm NC1} &=& 
  5.09\pm 0.63\times 10^6 \mbox{cm}^{-2}\mbox{s}^{-1},\\
\label{eq:NC2}
  R_{\rm SNO}^{\rm NC2} &=& 
  6.42\pm 1.67\times 10^6 \mbox{cm}^{-2}\mbox{s}^{-1}.
\end{eqnarray}
They are impossible to reconcile assuming the incident flux consists
entirely of electron neutrinos \cite{SNOPRL,SNOSK,SNONC}, and hence
some other particles must be contributing to the ES and NC event
rates.  Chronologically, the first detector to observe events caused
by these other particles was SK, but it was not possible to separate
these events from the $\nu_e$-induced events until the SNO CC data
became available. One can therefore say that the SNO measurement has
effectively reclassified SuperKamiokande as an appearance experiment.

The interpretation of the SNO results, however, is not unambiguous. 
While the ``excess'' NC and ES events are most often ascribed to 
muon (and tau) neutrinos, they can be also caused by muon (and tau)
\emph{anti}neutrinos.
The 
antineutrinos in question must be of the muon or tau types, since
electron antineutrinos would be easily identified through the reaction
$\bar\nu_e+p\rightarrow n+e^+$ \cite{TorrenteLujan1999}.
%which for
%$E_\nu=10$ MeV has a cross section two orders of magnitude greater
%than $\nu_e+e^-\rightarrow \nu_e+e^-$.
Both $\nu_{\mu,\tau}$ and $\bar\nu_{\mu,\tau}$ scatter on electrons
and deuterium nuclei through their neutral current (NC) interactions,
with similar cross sections. At $E_\nu\sim 10$ MeV,
\begin{eqnarray}
  \label{eq:simplecross}
  \int_{T_{\rm min}}^{T_{\rm max}}dT_e
      \frac{d\sigma(\bar\nu_{\mu,\tau}e^-)}{dT_e}/
  \int_{T_{\rm min}}^{T_{\rm max}}dT_e
      \frac{d\sigma(\nu_{\mu,\tau}e^-)}{dT_e} \sim 0.8,
\end{eqnarray}
where the integral over the electron recoil energy $T_e$ ranges from
$T_{\rm min}=5$ MeV to $T_{\rm max}=E_\nu/(1+m_e/2 E_\nu)$ MeV.
Similarly, using the latest available calculations of the
$\nu_x(\bar{\nu}_x)+d\rightarrow p+n+\nu_x(\bar{\nu}_x)$ cross
sections \cite{Kubodera} and integrating over the undistorted $^8$B
reaction spectrum, $f(E_\nu)$, \cite{8bspectrum}, one finds
\begin{eqnarray}
  \label{eq:NCcross}
  \int dE_{\nu} f(E_{\nu})
  \sigma(\bar\nu_{\mu,\tau}d)/
  \int dE_{\nu} f(E_\nu)
  \sigma(\nu_{\mu,\tau}d)\sim 0.95.
\end{eqnarray}

Assuming the $\nu_e\rightarrow\nu_{\mu,\tau}$ conversion mode,
Eqs.~(\ref{eq:CC}-\ref{eq:NC1}) imply that the
electron neutrino survival probability is $P_{ee}\simeq 0.35\pm 0.04$
($P_{ee}\simeq 0.27\pm 0.06$ if
Eqs.~(\ref{eq:CC}-\ref{eq:ESSK},\ref{eq:NC2}) are used). 
%and the inferred flux of all active neutrinos is 
%$f_{\rm tot}=5.4\pm 1.0\times 10^6$ cm$^{-2}$s$^{-1}$. 
If instead the $\nu_e\rightarrow\bar\nu_{\mu,\tau}$ mode is
assumed, the same data imply a somewhat higher flux of all active
neutrinos, namely, $\Phi_{\rm tot}=5.25\pm 0.66$ cm$^{-2}$ s$^{-1}$ if
one uses Eqs.~(\ref{eq:CC}-\ref{eq:NC1}) or $\Phi_{\rm tot}=6.64\pm
1.75$ cm$^{-2}$ s$^{-1}$ if one uses
Eqs.~(\ref{eq:CC}-\ref{eq:ESSK},\ref{eq:NC2}).  The corresponding
survival probabilities are $P_{ee}\simeq 0.34\pm 0.05$ or
$P_{ee}\simeq 0.27\pm 0.07$. Thus, both conversion modes are capable of
accounting for the measured event rates. Moreover, both yield the total
flux that is consistent with the Standard Solar Model (SSM), which
predicts $5.05_{-0.8}^{+1.0}\times 10^6$ cm$^{-2}$ s$^{-1}$
\cite{BP2000} or $5.93\pm 0.89\times 10^6$ cm$^{-2}$ s$^{-1}$
\cite{Bahcall_robust}, depending on the input value of the S17
parameter used. Therefore, in interpreting the experimental data both
possibilities should be considered.

The $\nu_e\rightarrow\nu_{\mu,\tau}$ conversion is predicted in the
flavor oscillation scenario\cite{MSW}, while
$\nu_e\rightarrow\bar\nu_{\mu,\tau}$ is predicted in the neutrino
spin-flavor flip (SFF) scenario
\cite{Cisneros,Okun1986,Akhmedov1988,LimMarciano,Raghavan}.  The
flavor oscillation scenario has been extensively studied over the
years. The goal of this paper is to reexamine the SFF scenario and to
point out the existence of a new solution.

In the SFF scenario, the
neutrino is postulated to couple to the electromagnetic field through
a magnetic moment, $\mu$, interaction. A neutrino propagating in the
solar magnetic field undergoes a spin rotation, and, if it is
a Majorana fermion, becomes an antineutrino, in the process
also changing its flavor. 
%Thus, the
%transition $\nu_e\rightarrow\bar\nu_{\mu,\tau}$ takes place.

Majorana fermions are most naturally described using the two-component
Weyl spinor notation. 
The coupling in question is given by a Lagrangian term
\begin{eqnarray}
  \label{eq:lagr}
  {\cal L}_{EM} = -\frac{1}{2}\mu_{a b} (\nu^\alpha)_a
  (\sigma^{\mu\nu})_{\alpha}^{\phantom{\alpha}\beta}
  (\nu_\beta)_b F_{\mu\nu} + {\rm h.c.},
%= \nonumber\\
%\mu_{a b} (\chi)_a
%  \vec{\sigma} (\nu)_b
%  (\vec{E}+i\vec{B}) + {\rm h.c.}
\end{eqnarray}
where
$(\sigma^{\mu\nu})_{\alpha}^{\phantom{\alpha}\beta}\equiv(\sigma_{\alpha\dot\alpha}^{\mu}\bar\sigma^{\dot\alpha\beta\:\nu}-\sigma_{\alpha\dot\alpha}^{\nu}\bar\sigma^{\dot\alpha\beta\:\mu})/2$,
$\sigma_{\alpha\dot\alpha}^{\mu}\equiv (1,\vec{\sigma})$,
$\bar\sigma^{\dot\alpha\beta\:\nu}\equiv (1,-\vec{\sigma})$. $\mu$ and
$\nu$ are Lorentz indices, $\alpha$ and $\beta$ are Weyl spinor
indices, and $a$ and $b$ are the flavor indices.  Since the spinors
anticommute, the $a=b$ terms in Eq.~(\ref{eq:lagr}) vanish
identically, and hence Majorana neutrinos cannot have magnetic
moments\cite{Valle1981}. They can, however, have \emph{transition}
($a\neq b$) moments which lead to a simultaneous spin and flavor
change. This provides a natural scenario for
$\nu_e\rightarrow\bar\nu_{\mu,\tau}$ transitions without also
producing $\bar\nu_e$ states. Flavor mixing, if present, must be small
to avoid a significant secondary
$\bar\nu_{\mu,\tau}\rightarrow\bar\nu_e$ conversion. Regarding the
naturalness of this assumption we note that out of the three mixing
angles in the leptonic sector the one measured with atmospheric
neutrinos is large and the one measured by reactor experiments is
small, so that {\it a priori} there is no natural value for the third
angle.

In general, the evolution of the neutrino state is governed by a
$6\times 6$ Hamiltonian matrix. For simplicity, we consider a
$4\times 4$ case. In the basis $(\nu_e,\nu_\mu,\bar\nu_e,\bar\nu_\mu)$,
the Hamiltonian is \cite{LimMarciano}
\begin{eqnarray}
\label{eq:H}
 H= \left(\begin{array}{cc}
 H_\nu & (B_x-i B_y)M^\dagger \\
 (B_x+i B_y)M &  H_{\bar\nu}
\end{array}\right),
\end{eqnarray}
where $B_{x,y}$ are the transverse components of the magnetic field
and the $2\times 2$ submatrices are given by
\begin{eqnarray}
\label{eq:H1}
M&=&\left(\begin{array}{cc}
  0& -\mu_{e\mu} \\
 \mu_{e\mu} &  0
\end{array}\right),\\
\label{eq:H2}
 H_{\nu}&=&\left(\begin{array}{cc}
  -\Delta\cos 2\theta + A_e& \Delta\sin 2\theta \\
 \Delta\sin 2\theta &  \Delta\cos 2\theta + A_\mu
\end{array}\right),\\
\label{eq:H3}
 H_{\bar\nu}&=&\left(\begin{array}{cc}
  -\Delta\cos 2\theta - A_e& \Delta\sin 2\theta \\
 \Delta\sin 2\theta &  \Delta\cos 2\theta - A_\mu
\end{array}\right).
\end{eqnarray}
Here $\Delta\equiv\Delta m^2/4 E_\nu$, 
$A_e\equiv\sqrt{2}G_F(n_e-n_n/2)$ and 
$A_\mu\equiv\sqrt{2}G_F (-n_n/2)$. 
$\Delta m^2$ is the neutrino mass-squared splitting, $E_\nu$ is its
energy, and $n_e$ and $n_n$ are the electron and neutron number
densities. 

As will be described later,
to obtain a significant $\nu_e\rightarrow\bar\nu_{\mu,\tau}$
conversion in the Sun requires $\mu_{e\mu} \sim (10^{-12}-10^{-10})
\mu_B$ ($\mu_B\equiv e/2 m_e$ is the Bohr magneton). The transition
moment values in this range are below the direct laboratory bounds
derived from the analyses of the $\nu+e\rightarrow\nu+e$
\cite{Krakauer} and $\bar\nu+e\rightarrow\bar\nu+e$ \cite{Derbin}
scattering data.  Stronger bounds on the transition moment have been
claimed from the analysis of various astrophysical processes, most
notably from the study of the red giant populations in globular
clusters \cite{Raffeltbound}.  The idea is that a sufficiently large
transition moment would provide an additional cooling
mechanism and change the red giant core mass at helium flash beyond
what is observationally allowed. The bound stated in
\cite{Raffeltbound} is $\lesssim 3\times 10^{-12}\mu_B$. Recently,
there has been some disagreement between the stellar models and also a
systematic shift to longer distances for clusters \cite{Achim}, and it
would be very interesting to see an updated bound.

The compilation of various bounds on the size of the transition moment
and the corresponding references can be found in \cite{pdg}.

\section{Magnetic field in the Convective Zone}

%\textit{Magnetic field in the Convective Zone.}---
The neutrino spin-flavor flip can occur either in
the convective zone (CZ) $(0.71 R_\odot\lesssim r<R_\odot)$ or in
the radiative zone (RZ) deeper in the solar interior. The two
possibilities are physically quite different and require separate
treatments. First, we consider the CZ case.

While the magnetic field in the inner part of the CZ cannot yet be measured
directly, its basic properties can be inferred from the
field at the solar surface, and it is widely believed that
large scale field structures exist at some depth in the CZ
\cite{Parker}. Recent magnetohydrodynamic dynamo models argue that the
field generation occurs in the shear layer near the bottom of the CZ.
%(sometimes referred to as the ``interface layer''). 
These models predict field values as large as 100 kG \cite{Parker},
which significantly exceeds the turbulent equipartition value of
$B\sim \rho^{1/6}L_\odot^{1/3}r^{-2/3}\sim 10^4$ G
\footnote{This estimate is obtained by estimating the convective energy flux as
$L_{\odot } \sim 4\pi r^2 v\rho v^2/2$, where $v$ is the
turbulent velocity,  and assuming equipartition
$\rho v^2/2 \sim B^2/8\pi $.}.  The helioseismological data provide an upper
limit of 300 kG on the magnitude of this field \cite{antia}.

The field strength necessary for significant
$\nu_e\rightarrow\bar\nu_{\mu,\tau}$ conversion can be estimated as
$\mu_{e\mu}B\Delta l\sim 1$, where $\Delta l$ is the thickness of the
magnetic field.  One finds
\begin{equation}\label{eq:Best}
  (\mu_{e\mu}/\mu_B) (B/1 \mbox{ kG})
  (\Delta l/0.1R_\odot)\sim 5\times 10^{-10},
\end{equation}
so that for $\mu_{e\mu}\sim 10^{-11}\mu_B$ and $\Delta l \sim 0.1 R_\odot$ 
the value $B\sim 50$ kG is required.  Although the surface field,
which reaches several kG in sunspots, is too weak to affect the
evolution of the solar neutrinos, the field near the bottom of the CZ
might have the right strength (provided $\mu_{e\mu}\sim 10^{-11}\mu_B$). 
This observation, and the natural expectation that the field in the
interior of the CZ changes with the solar cycle, were the reasons the
convective zone SFF were originally suggested as an explanation for an
apparent anticorrelation between the Homestake event rate and the
number of sunspots \cite{Okun1986}.

Over time, however, a paradigm shift has occurred. The idea of using this
mechanism to explain variations of the neutrino rates was abandoned,
especially in view of the SK data, which showed no seasonal or yearly
variation, except for a hint of the expected $1/r^2$ flux modulation.
Instead, recent analyses \cite{CZfits} assume that the magnetic field
in the CZ is somehow time independent, an assumption that we find untenable.
%No model of the solar magnetic field predicts the
%surface field undergoing periodic 11 year reversals while the large
%scale field below is constant. 
The presence of a strong constant magnetic field inside the CZ would,
on the one hand, almost certainly be revealed at the surface by
convective mixing and, on the other hand, cause an asymmetry in the
solar cycle. The strong toroidal field that is believed to exist in
the overshoot layer, where convective mixing is small and the shear is
large, should also be variable because it is generated from a variable
radial field. It is therefore implausible that the SFF process in the
CZ could cause a $\sim70$\% depletion of the $^8$B neutrino flux,
while producing no observable variations at SK during more than four
years.  This time span covers about half of the solar cycle, with an
average sunspot number $\sim 10$ in 1996, and $\sim 120$ in 2000
\cite{sunspots}.

In addition to the year-to-year variations, one also expects
characteristic semiannual variations \cite{Okun1986}, for the
following reason. The field on the surface of the Sun changes
direction across the solar equator and vanishes at the equator (where
indeed no sunspots are observed). The transition region, where the
field is small, has the size $\sim 7\times 10^9$ cm
\cite{Okun1986,BahcallBook}.  The orbit of the Earth makes a $7^\circ$
angle with the plane of the solar equator and, since 95\% of the $^8$B
neutrinos are produced in the region $r<0.09 R_\odot$ \cite{BP2000}, a
significant fraction (88\%) of them passes through the
transition region when the Earth crosses the plane of the solar
equator. On the other hand, in March and September, when the Earth is
farthest from the solar equatorial plane, this fraction is as low as
1\%. Thus, large semiannual variations are expected at
Super-Kamiokande and are not observed \cite{SK2001}.

We therefore regard the absence of variations in the SK data as 
strong evidence against SFF in the CZ.

\section{Relic field in the RZ}

%\textit{Relic field in the RZ.}---
We now turn to the possibility that the SFF process could
occur in the radiative zone of the Sun. Unlike the convective zone,
the radiative zone is not continuously mixed and rotates as a
solid body. Therefore, it could in principle support a ``frozen''
magnetic field configuration.

The properties of magnetic fields in the solar interior were
investigated by Cowling \cite{cowling}, and by Bahcall and Ulrich
\cite{bahcall_ulrich}. These authors consider poloidal magnetic
fields. We have redone the calculation \cite{magnapj} using a modern
solar model (BP2000), and, more importantly, assuming that the field
is toroidal. A strong poloidal field would penetrate the convective
zone and would either dissipate over the solar lifetime, or be
observable. We also assumed axisymmetry, which is physically
plausible. Large asymmetry would lead to observable 27-day variations
of the neutrino rate due to solar rotation.

\begin{figure}[t]
  \begin{center}
    \includegraphics[angle=0, width=.4\textwidth]{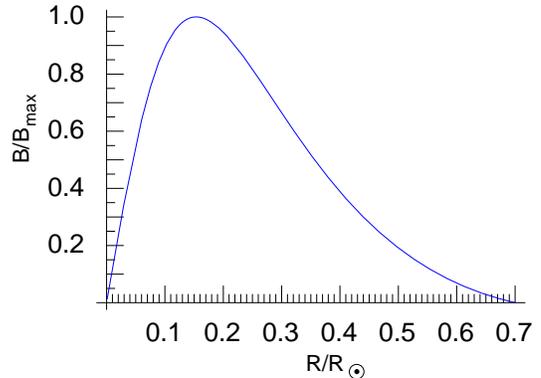}
    \caption{The radial dependence of the field strength for the
      lowest toroidal mode in the RZ of the BP2000 SSM. }
    \label{fig:prfl}
  \end{center}
\end{figure}

For an axisymmetric toroidal field, the magnetic diffusion equation
(describing Ohmic decay) was solved numerically. The eigenmodes are of
the form $B=e^{-t/\tau_{l,n}}F_{l,n}(r)P_l^1(\cos \Theta )$, where
$l=1,2,..$, $n=1,2,..$, $P_l^1(\cos\Theta)$ is the associated Legendre
polynomial and $F(r)$ is determined numerically. The $l=1$, $n=1$ mode
has the longest lifetime, $\tau _{1,1}= 24$ Gyr; its radial profile is shown in
Fig.~\ref{fig:prfl}. For higher modes, $\tau _{2,1}= 13$ Gyr,
$\tau _{1,2}= 10$ Gyr, ... For
comparison, the solar age 
is thought to be 4.6 Gyr \cite{BP2000}. Therefore, a
toroidal field of complex spatial structure can exist in the RZ of the
Sun. 

Observations provide upper bounds on the allowed strength of this field
 \cite{magnapj}. Neutrino fluxes and helioseismological
measurements of the sound speed 
both give $B\lesssim 10^2$ MG. Possible residual (non-rotational)
oblateness and helioseismological measurements of the frequency
splittings provide stronger bounds, $B\lesssim 7$ MG \cite{magnapj}.

We next consider the neutrino evolution in the presence of this relic
field and show that with appropriate parameter choice there exists a
fit to all available solar neutrino data. 

For simplicity, we assume no flavor mixing and that only the longest
living eigenmode is present. This is sufficient to demonstrate the
existence of the solution. The neutrino
evolution equations then reduce to a pair of coupled equations,
\begin{eqnarray}
  \label{eq:nuevolution}
  i\partial_l\psi_e &=&
  (A_e(r)-\Delta)\psi_e+\mu_{e\mu}B_\perp(l)\psi_{\bar\mu}, 
  \nonumber\\
  i\partial_l\psi_{\bar\mu} &=&
  \mu_{e\mu}B_\perp(l)\psi_e+(\Delta-A_\mu(r))\psi_{\bar\mu}.
\end{eqnarray}
Here $l$ is the distance along the neutrino trajectory and $B_\perp$
is the transverse component of the magnetic field. There are two
free parameters, $\Delta m^2$ and the product of the neutrino magnetic
moment $\mu_{e\mu}$ and the normalization of the
magnetic field $B_{\max}$. For definitiveness, we set 
$\mu_{e\mu}=10^{-11}$ $\mu_B$. Later we will address the possibility
of having different values of $\mu_{e\mu}$. 

\begin{figure}[t]
  \begin{center}
    \includegraphics[angle=0, width=.41\textwidth]{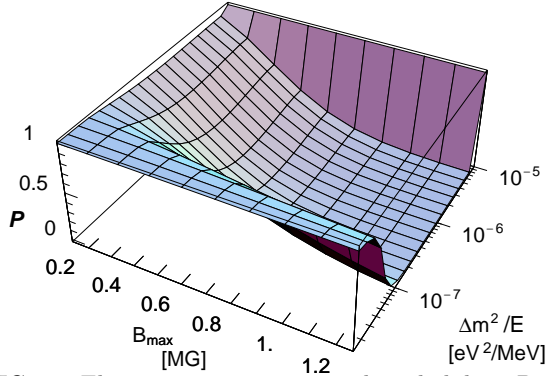}
    \caption{Electron neutrino survival
      probability $P_{ee}$ as a function of $B_{\rm max}$ and $\Delta
      m^2/E_\nu$, for $\mu_{e\mu}=10^{-11}$ $\mu_B$.}
    \label{fig:scan}
  \end{center}
\end{figure}

\begin{figure}[t]
  \begin{center}
    \includegraphics[angle=0, width=.45\textwidth]{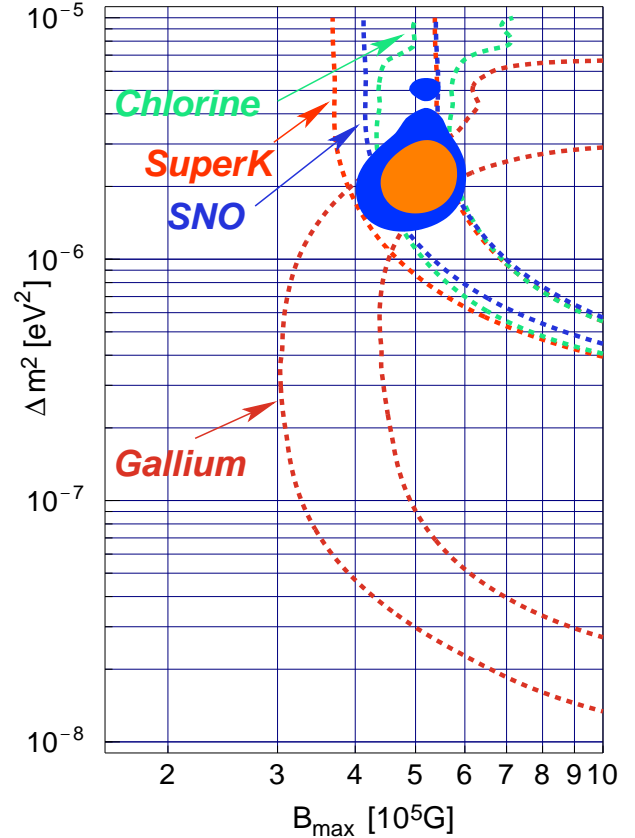}
    \caption{Regions in the $(\Delta m^2, B_{\rm max})$ parameter
      space where the rates of various solar neutrino experiments are
      reproduced.}
    \label{fig:fit}
  \end{center}
\end{figure}

\begin{figure}[t]
  \begin{center}
    \includegraphics[angle=0, width=.45\textwidth]{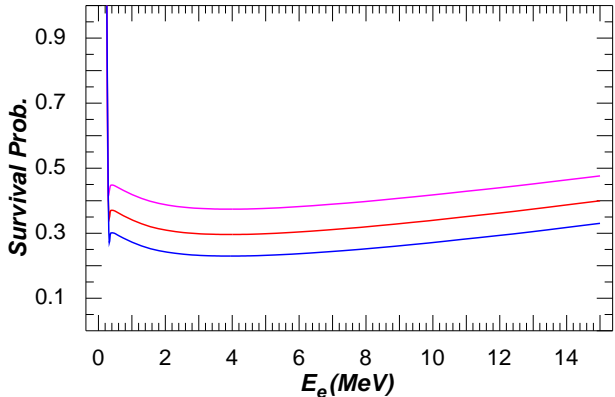}
    \caption{Survival probabilities for an electron neutrino born in
      the center of the Sun for three values of $B_{\rm max}$: 0.45
      MG, 0.5 MG, and 0.55 MG (top to bottom). $\Delta m^2=2.5\times
      10^{-6}$ eV$^2$ was chosen.}
    \label{fig:PvsEnu}
  \end{center}
\end{figure}

Fig.~\ref{fig:scan} shows the survival probability $P_{ee}$ for
electron neutrinos produced in the center of the Sun, as a function of 
$B_{\rm max}$ and $\Delta m^2/E_\nu$. The probability was obtained by
solving the evolution equations (\ref{eq:nuevolution}) numerically on
a grid of points.

The
behavior of $P_{ee}$ can be given a simple analytical interpretation.
The situation is completely analogous to the classical case of the
small angle MSW effect \cite{LimMarciano,Akhmedov1988}. The SFF
process takes place in a thin layer around $r_{\rm res}$ given by the
condition $A_e(r_{\rm res})+A_\mu(r_{\rm res})=2\Delta$.  For 
$\Delta m^2/E_\nu\gtrsim 10^{-5}$ eV$^2$/MeV, the resonance never
occurs and therefore $P_{ee}\rightarrow 1$. For 
$\Delta m^2/E_\nu\lesssim 4\times 10^{-8}$ eV$^2$/MeV, the resonance occurs
close to the edge of the RZ where the field is small and hence once
again $P_{ee}\rightarrow 1$ 
(or outside in the CZ, in which case $P_{ee}$ varies with time). In
the intermediate region, the conversion efficiency depends on the
value of $\mu_{e\mu}B_\perp$ at $r=r_{\rm res}$: for $\mu_{e\mu}B_\perp$
$\gtrsim 0.7\times 10^{-11}$ MG$\times\mu_B$ the conversion is adiabatic and
$P_{ee}\rightarrow 0$.

Notice that for this description it is crucial that everywhere except
in the thin resonance layer the off-diagonal term $\mu_{e\mu}B_\perp$
is much smaller than the diagonal splitting
$(A_e(r)+A_\mu(r))/2-\Delta$. This allows one to ignore the fact that
the off-diagonal coupling varies along the neutrino trajectory and use
a small angle MSW formula for the level crossing probability
\begin{equation}
  \label{eq:Pc}
  P_c=\exp\left(-\pi
    \frac{2(\mu_{e\mu} B_\perp(l))^2}{|d(A_e+A_\mu)/dl|}\Big
    |_{l=l_{\rm_{res}}}\right).
\end{equation}
The complications related to the large angle MSW effect
\cite{largeangleMSW} are likewise avoided. The neutrino survival
probability is given by $P_{ee}=(1+(1-2 P_c)\cos 2\theta_\odot)/2$,
where
$\cos 2\theta_\odot=(2\Delta-A_e-A_\mu)[(2\Delta-A_e-A_\mu)^2+(2\mu_{e\mu}B_\perp)^2]^{-1/2}$
is computed at the production point.

We performed a fit to all available solar neutrino data.
In our computations, we used the fluxes of the \textsl{pp}, $^7$Be,
$^8$B, $^{13}$N, $^{15}$O, and \textsl{pep} neutrinos, as predicted by
the BP2000 solar model \cite{BP2000}.
The data to be fitted include the average
rates of the GALLEX/GNO \cite{GALLEX} and SAGE \cite{SAGE}
experiments,
\begin{eqnarray}
  \label{eq:GArates}
  R_{\rm GA}=74.7\pm 5.1\; {\rm SNU},
\end{eqnarray}
the rate of the Homestake experiment \cite{Homestake},
\begin{eqnarray}
  \label{eq:CHLrates}
  R_{\rm CHL}=2.56\pm 0.23\; {\rm SNU},
\end{eqnarray}
the CC and NC rates of the SNO experiment as given in
Eqs.~(\ref{eq:CC},\ref{eq:NC1}), and the rate and the energy
spectrum of the SK experiment \cite{SK2001}.

For a given production point in the solar core, the neutrino survival
probability is determined according to Eqs.~(\ref{eq:nuevolution}). To
obtain the total flux of the neutrinos of a given energy, one needs to
integrate over the neutrino production regions. The problem is
technically quite nontrivial, because of the 3-dimensional structure
of the field and because only the field component transverse to the
neutrino trajectory enters the evolution
equations.\footnote{Fortunately, in the case of the lowest field
  eigenmode a considerable simplification occurs because of the
  $\sin\Theta$ dependence of the field on the polar angle. Still, a
  2-dimensional numerical integration is required.} The correct
integration is absolutely essential for computing the converted
fraction of the low energy \textsl{pp} neutrinos, which have a broad
production region and for which the resonance occurs close to, and
even inside, that region.

Fig.~\ref{fig:fit} shows regions of the parameter space where the
event rates predicted for various experiments agree with the
corresponding measured rates. The dashed lines delineate the bands
where agreement is reached with a particular experiment. The
intersection region, where a good fit to all rates is obtained, is
characterized by $0.45$ MG$\lesssim B_{\rm max}\lesssim 0.55$ MG and
$1.5\times 10^{-6}$ eV$^2\lesssim\Delta m^2\lesssim 3\times 10^{-6}$
eV$^2$. It represents our new solution to the solar neutrino deficit.

As an illustration, we list below our predictions for the point
$\Delta m^2=2.5\times 10^{-6}$ eV$^2$ and $B_{\rm max}=0.5$ MG.  We
find that for this point the $^8$B flux is suppressed by a factor of
0.34, the \textsl{pp} flux -- by a factor of 0.86, and the $^7$Be flux
-- by a factor of 0.36.  The $^7$Be suppression factor implies the
effective Borexino flux suppression of $0.44$, which is
below $0.65_{-0.12}^{+0.14}$ expected for the LMA solution
\cite{Bahcall_robust}.  The predicted rates are $80.6\pm 2.6$ SNU for
the Ga experiments and $2.69\pm 0.31$ SNU for the Cl experiment. The
corresponding $\chi^2$ for the rates fit is 3.4 for 4 d.o.f.  

In Fig.~\ref{fig:PvsEnu} we show the survival probability for a
neutrino produced in the center of the Sun as a function of the
neutrino energy, taking $\Delta m^2=2.5\times 10^{-6}$ eV$^2$. The
three curves correspond to three different normalizations of the
magnetic field eigenmode, $B_{\rm max}=0.45,0.5,0.55$ MG. We caution
the reader that the flux of the \textsl{pp} neutrinos cannot be easily
read off from this plot, since, as already mentioned, to compute it
one has to appropriately integrate over the production region.

\begin{figure}[t]
  \begin{center}
    \includegraphics[angle=0, width=.43\textwidth]{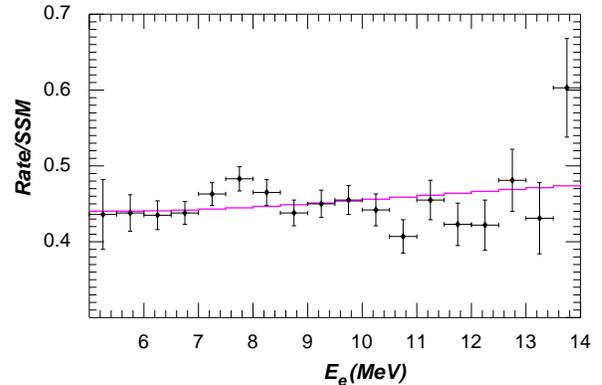}
    \caption{Predicted SuperKamiokande recoil electron energy spectrum
      for $B_{\rm max}=0.50$ MG and $\Delta m^2=2.5\times 10^{-6}$
      eV$^2$. }
    \label{fig:spectrum}
  \end{center}
\end{figure}

The survival probability is not strictly constant for high energy
$^8$B neutrinos, however, its variation with energy is rather small.
The recoil electron energy spectrum expected at SK is shown in
Fig.~\ref{fig:spectrum}. The value of the $\chi^2$ is 24 for 19 d. o.
f. The spectrum becomes flatter as one increases $\Delta m^2$.

%The quality of the overall fit is therefore
%quite good.

It is worth stressing that the value of the field that gives a good
fit trivially scales with the value of the transition moment according
to $B'_{\rm max}=B_{\rm max} \times 10^{-11} \mu_B/\mu_{e\mu}$. In
particular, if one chooses $\mu_{e\mu}=3\times 10^{-12}\mu_B$ in order
to satisfy the astrophysical bound given in \cite{Raffeltbound}, one
can choose $B'_{\rm max}\sim 1.5-2$ MG, which is still perfectly
acceptable, as shown in \cite{magnapj}. No property of the field, such
as its lifetime, {\it etc.}, is affected by such a rescaling.

Our scenario may be convincingly tested (and perhaps excluded) by the
KamLAND experiment, should it establish flavor oscillations with the
LMA parameters, or by the Borexino experiment, should it observe large
day-night variations, as predicted for the LOW solution (SFF predicts
no variation).

There are three potential sources of time variations in our scenario.
First, the effect of the time varying CZ field
 can be shown to be negligibly small, assuming $B_{CZ}\lesssim 100$ kG
and $B_{RZ}$ and $\Delta m^2$ in the range of the fit.
Second,
if the field is not axially symmetric, one expects 27-day
variations.  Third, if, in addition to $l=1$ modes, the $l=2$ modes
($P_2^1(\cos\Theta)\propto \sin 2\Theta$) are also present, the flux
may exhibit annual variations, for the same reason that semiannual
variations are expected for the CZ SFF. These variations have
extrema in March and September, and hence are distinguishable from the
variations due to the eccentricity of the Earth's orbit. Calculations
show that for the points in the allowed region a relative change in
the SK rate is $\delta R/R\sim 2\delta B/B$.  Hence, even if the $l=2$
and $l=1$ modes have comparable amplitudes, the rate variation is
expected to be of the order of a few percent, in contrast to SFF in
the CZ, in which case it is expected to be much larger.  The current data
provide a weak upper bound on the strength of the $l=2$ modes. 
Both the 27-day and the annual variations described above 
constitute smoking gun signatures of the RZ SFF mechanism.

In summary, our scenario provides a 
%good 
fit to all
available solar neutrino data, is compatible with all existing
constraints, and may be tested in the very near future. 

\acknowledgements

We would like to 
thank John Bahcall, Sarbani Basu, Axel Brandenburg,
Gia Dvali, and Cecilia Lunardini for 
useful discussions. A. F. was supported by the W.~M.~Keck Foundation.

\end{document}